# Selecting active matter according to motility in an acoustofluidic setup: Self-propelled particles and sperm cells

Vyacheslav R. Misko[a,b,*], Larysa Baraban[c], Denys Makarov[d], Tao Huang[d], Pierre Gelin[a], Ileana Mateizel[e], Koen Wouters[e], Neelke De Munck[e], Franco Nori[b,f,g] and Wim De Malsche[a]

[a] *µFlow group, Department of Chemical Engineering, Vrije Universiteit Brussel, Pleinlaan 2, 1050 Brussels, Belgium.*
[b] *Theoretical Quantum Physics Laboratory, Cluster for Pioneering Research, RIKEN, Wako-shi, Saitama 351-0198, Japan.*
[c] *Helmholtz-Zentrum Dresden-Rossendorf e.V., Institute of Radiopharmaceutical Cancer Research, Bautzner Landstrasse 400, 01328 Dresden, Germany.*
[d] *Helmholtz-Zentrum Dresden-Rossendorf e.V., Institute of Ion Beam Physics and Materials Research, Bautzner Landstrasse 400, 01328 Dresden, Germany.*
[e] *Brussels IVF - Center for Reproductive Medicine, UZ Brussel, Laarbeeklaan 101, 1090 Brussels (Jette), Belgium.*
[f] *Quantum Computing Center, RIKEN, Wako-shi, Saitama, 351-0198, Japan.*
[g] *Physics Department, University of Michigan, Ann Arbor, Michigan 48109-1040, USA.*

Active systems – including sperm cells, living organisms like bacteria, fish, birds, or active soft matter systems like synthetic "microswimmers" – are characterized by motility, i.e., the ability to propel using their own "engine". Motility is the key feature that distinguishes active systems from passive or externally driven systems. In a large ensemble, motility of individual species can vary in a wide range. Selecting active species according to their motility represents an exciting and challenging problem. We propose a new method for selecting active species based on their motility using an acoustofluidic setup where highly motile species escape from the acoustic trap. This is demonstrated in simulations and in experiments with self-propelled Janus particles and human sperm. The immediate application of this method is selecting highly motile sperm for medically assisted reproduction (MAR). Due to the tunable acoustic trap, the proposed method is more flexible than the existing passive microfluidic methods. The proposed selection method based on motility can also be applied to other active systems that require selecting highly motile species or removing immotile species.

## 1. Introduction

To fertilize the human oocyte, the sperm must be able to travel through the female reproductive tract. Therefore, sperm motility (graded as A: progressive motile with high velocity, B: progressive motile with lower velocity, C: non-progressive motile, D: immotile) plays a crucial role in couple's reproductive potential. A decrease in progressive sperm motility will decrease the chance of fertilization, and consequently the chance to achieve a pregnancy. In couples where the partner presents a reduced sperm motility, medically assisted reproduction (MAR) can provide the means to select motile sperm that can be further used for either Intra Uterine Insemination, In Vitro Fertilization, or for Intracytoplasmic Sperm Injection. Commonly used sperm preparation techniques are swim-up and density gradient centrifugation [1]. Although these procedures increase the number of progressive motile sperm, they differ from the physiological way of sperm selection. To model stringent sperm selection processes that takes place in the female reproductive tract, more advanced selection procedures have been developed based on the sophisticated morphological assessment, electrical charge, and molecular binding. The latter two principles mimic in part female selection mechanisms [1].

Various microfluidic setups have been proposed to select progressive motile sperm cells. A passively driven self-contained integrated microfluidic device for separation of motile sperm has been presented [2] that isolates motile sperm based on its ability to cross streamlines in a laminar fluid stream [2, 3, 4]. Similarly, a variety of sperm-selecting devices have been proposed, including a rather sophisticated "sperm syringe" [5] and other devices [6], employing the same passive selection principle as the early works [2, 3, 4, 7]. In an attempt to model the environment in the bovine reproductive tract, where sperm is guided through the cervix by entering microgrooves in the walls of the endocervical canal, a device was proposed modeling the grooves of the cervix [8] allowing separation of motile from immotile bull sperm. More sophisticated microfluidic techniques imply the analysis of the individual properties of sperm cells and selecting the most active cells on that basis. Thus, a device called "spermometer" was proposed [9, 10] equipped with a sperm trap between microfluidic channels where the intensity of the beats of the sperm cell flagellum was characterized by measuring the impedance of the sperm trap with a sperm cell inside. The amplitude of the beats was converted into the amplitude of the electric signal. This sperm-selecting device included a feed-back circuit controlling a gate that opened when the sperm cells with the highest amplitude of flagellum beats were detected and closed otherwise [9, 10].

Other microfluidic methods of motile sperm selection do not require external flows to drive sperm. Instead, they use own activity of motile sperm cells to select them from immotile cells and debris. The mechanism of selection in this class of devices is related to the ability of motile sperm cells to be trapped and accumulate in the corners of asymmetric obstacles [11] or move in the preferable direction through microchannels [12, 13, 14] or in an array of asymmetric obstacles [15]. The latter effect can be referred to as self-rectification of active motion in asymmetric environment [16]. It should be noted that self-rectification of active motion, despite general similarities, has an essential difference with rectification in driven dynamics [17]: it does not require any external correlated noise but occurs due to the finite persistence length of the motion of self-propelled particles (or cells) [16]. Related numerical studies discuss a selective escape of microswimmers from circular domain, depending on the motility [18, 19].

The above techniques imply that progressive motile sperm can be selected and separated from all the available sperm and then used for MAR. However, it can occur that the selected sperm does not have the capacity of successful fertilization, due to reduced motility. In this case, it would be desirable to *enhance* their motility. Following this strategy, advanced methods were proposed that employ high-activity guest microswimmers (other than sperm) interacting with sperm cells to enhance their motility. Thus, sperm carrying micromotors were proposed [20] where single immotile live sperm cells were captured, transported, and released in fluidic channels, and delivered to the oocyte for fertilization. An alternative method is based on the effect called "motility transfer" [21] where a more active guest species in a binary mixture of active swimmers transfers its motility to a less active host species. It was proposed to use high-motility artificial microswimmers, i.e., bio-compatible catalytic Janus particles [22], whose activity can be controlled. The proposed "motility transfer" technique [21] has advantages over other similar proposals, as it is substantially less damaging, and it does not require the fast guest swimmers to localize and trap individual sperm cells.

In this work, we demonstrate a new principle of selecting motile active matter, e.g., progressive motile sperm, using acoustofluidic setup. The idea behind the proposed method is that the flow of sperm (or motile particles) and debris (immotile particles) can be focused inside a continuous fluid flow as recently demonstrated for particles [23]. The most active sperm can escape the focused flow, due to their high motility, similarly to the escape from the flow in passively-driven sperm separators [2, 3, 4], while the rest – including immotile sperm and debris – remain focused. The acoustic trapping has been demonstrated for sperm cells [24, 25], other (immotile) living cells [26, 27], bacteria [28], micro- [23] and nanoparticles [28]. However, the previous studies on sperm cells only performed selective trapping based on size, allowing for instance separation of sperm from debris [22, 23], but did not address separation based on motility.

In our study, we focus on separation of sperm according to their *motility*, i.e., on the ability of highly motile sperm cells (i.e., progressive motile sperm with high velocity) to escape from the acoustic trap. Thanks to the acoustic trap with a tunable strength, the proposed method provides much more flexibility in selecting sperm cells based on their motility as compared to passive methods where sperm cells are focused by the flow itself. Indeed, in passive devices [2-4] all the species, motile and immotile, can escape the flow either due to active propulsion or due to diffusion (with different escape times). In our proposed method, passive (or less motile) species are selectively *trapped* by the acoustic potential and therefore cannot escape, while active swimmers like self-propelled particles or highly motile sperm cells possess sufficient energy to overcome the potential energy barrier created by acoustic wave. The height of this potential energy barrier is tuned by the applied acoustic power from zero (passive regime) to high values when all the species can be trapped. Therefore, this delivers high flexibility of the proposed method in selecting, e.g., only the most motile species or all motile species and separating them from immotile species. Furthermore, our method is non-contact, which significantly reduces the danger of sperm cell damage. The principle of sperm cell selection and separation proposed in this work has been first demonstrated in numerical simulations and verified in experiments with active Janus particles (which serve as a model system of living sperm cells) and finally in experiments with human sperm cells. We emphasize however that these experiments have demonstrated the principle and possibility to select and separate active matter based on its motility while the design and analysis of devices for practical applications (e.g., for sperm separation for MAR) will require further efforts which is beyond the scope of the current study.

## 2. Results and Discussion

The principle of the proposed acoustic separation method is illustrated in Fig. 1(a). The complete sperm sample including motile and immotile sperm and debris (or other motile and immotile species, e.g., artificial self-propelled particles and passive beads), is injected in the channel where they follow the fluid flow. In the region of the acoustic actuator, all the cells and particles experience the pressure from the radiation force and in this way become focused in the middle of the channel. High motility sperm cells can escape from the focused flow while immotile sperm and other inactive particles cannot, and they follow the central flow. At the end of the device, high motility sperm cells can be collected through the side channels, separately from the rest of the sample that is collected through the central channel. Depending on the strength of the acoustic focusing, motile sperm cells can either be focused together with other immotile species (Fig. 1b), if the focusing potential is strong enough, or they can escape the trap already in the region of the acoustic actuator, if the focusing potential is insufficient to trap highly motile sperm cells. In either case they move outside the central focused flow and finally can be collected through the side channels.

The physical mechanism behind the selection of motile sperm cells (or self-propelled particles) based on their motility can be understood as follows. All the species, motile and immotile, are trapped by the focusing potential, and only motile species escape while immotile remain trapped. Immotile species execute Brownian diffusion (which is appreciable for particles with sizes below 1 μm). They experience the radiation force directed towards the centerline of the channel. This force prevents their Brownian diffusion out of the trap. Active species execute active Brownian diffusion (or run-and-tumble type motion in case of sperm cells) that is characterized by a finite persistence length, i.e., a straight part of the trajectory of their motion. In general, the direction of motion of active species is random: it could be along the channel or perpendicular to the direction of the channel. When an active swimmer moves away from the center, it can overcome the radiation force that pushes it towards the centerline of the channel. Recall that in case of overdamped dynamics the velocity is proportional to the force, i.e., the higher the self-velocity the higher potential trap the particle can overcome. This essential difference in motion of motile and immotile species allows the former to escape the acoustic trap. Furthermore, it was shown [29] that Pt-covered Janus particles driven by a flow in a microchannel, drift *across* the channel, i.e., across the flow streamlines, from the centerline towards the channel walls. This effect additionally contributes to the escape of Janus particles from the trap.

## 2.1. Numerical experiment with motile self-propelled particles

It is common to consider artificial "microswimmers", i.e., synthetic self-propelled particles like Janus particles (JP), as an active soft-matter system mimicking the motion of motile sperm cells or living organisms such as bacteria, or even macroscopic self-propelled "motors" as swimming fish, flying birds, under conditions far from the thermodynamic equilibrium (see, e.g., [30]). Here we explore this striking analogy and analyze the motion of sperm cells in the acoustic separation device assuming that, the formalism developed for artificial self-propelled particles and successfully employed for the description of a great variety of phenomena [16, 21, 22, 30-37], fairly describes the motion of motile sperm. We therefore abstract from the details of sperm morphology and propulsion mechanism and focus on the active diffusion of sperm cells exploring the analogy between the motion of sperm and artificial self-propelled particles [38, 39].

The equations of motion of self-propelled particles and the details of the simulation method are presented in the "Methods" section. It is worth noting that in our model, diffusion of passive and active species, as well as active diffusion of self-propelled particles, is taken into account. In this way, passive species have the ability to leave the central flow over time, due to dispersion. The interaction among the species is modelled by the elastic core-to-core interaction. We emphasize that in the simulations we aim to reveal the effect, and not to reproduce any specific fixed experimental conditions as these may vary from system to system. Therefore, we use dimensionless parameters normalized on the well-defined units: the size of the particle (unit length) and the flow velocity (unit velocity). This approach allows us to reveal the *optimal* relations between the parameters required to observe the selection of self-propelled particles based on their motility. These results can then be easily mapped on a specific system.

The results of the simulations are presented in Fig. 1(b-c) and Fig. 2. First, we consider a two-component binary mixture of active JP and passive particles. Initially, all the particles are injected and homogeneously distributed along the cross-section of the channel where they follow the fluid flow. Inside the region of action of the acoustic potential, the flow of particles becomes focused. If the focusing potential is strong enough, then all the particles become trapped (Fig. 1(b)), and at a "detector" ($x$ = 15) motile and immotile species can be detected separately: immotile species follow the central focused flow while motile species run away from the central flow and can be collected at side detectors. In this numerical experiment the flow velocity (for simplicity, it is taken homogeneous across the entire channel although it can be chosen, e.g., having a parabolic profile, which does not influence the results) is taken twice the value of the self-propulsion velocity of the active species. In Fig. 1(c), the results are presented for the case when the trapping potential is reduced to a half of its value as compared to the situation shown in Fig. 1(b).

A somewhat more complex situation is presented in Fig. 2 when motile species of different motility are present in combination with immotile species. Thus, we assume that the mixture contains three sorts of species: highly motile sperm with self-velocity $v$ = 2$v_{flow}$, where $v_{flow}$ is the fluid flow velocity (that can be easily controlled in the experiment), moderate-motility sperm with self-velocity $v$ = $v_{flow}$, and immotile sperm (or debris) with $v$ = 0. When the trapping acoustic potential is strong enough, all the species are focused inside the region of the action of the potential (Fig. 2(a)). Then motile species escape the focused flow, and this escape is much stronger for the highly motile sperm. Note that these highly motile species can even move against the flow (in the same way as real motile sperm cells in the experiment) which can be seen in the figure as loops in their trajectories. Now the motile species can be separately collected, e.g., at position x = 15 (Fig. 2(a)). Note that, if we are interested in collecting only the most active, highly motile sperm (for the needs of the assisted fertilization), then the collecting chambers should be placed at a larger distance from the central line of the channel as can be seen from Fig. 2(a). A similar behavior is observed when the trapping acoustic potential is reduced (Fig. 2(b)).

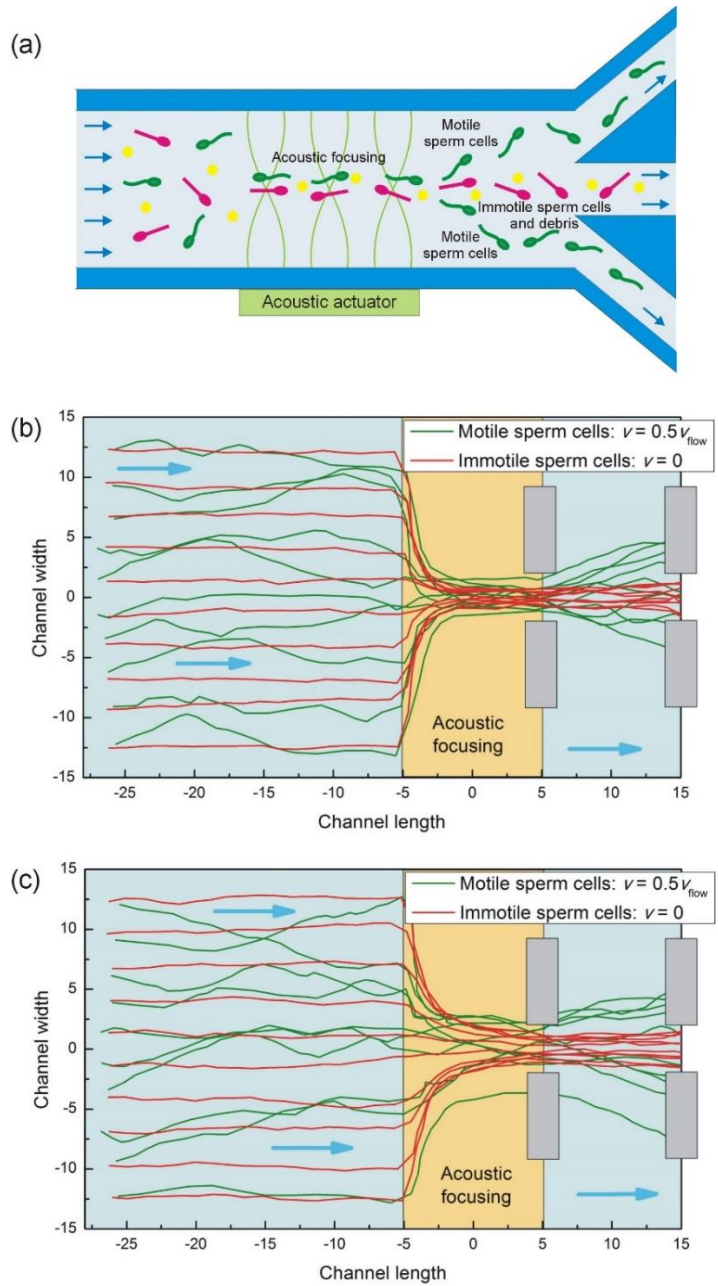

**Figure 1.** (a) Principle of acoustic sperm (or other motile species like self-propelled Janus particles) separation according to their motility. The sample, including sperm (motile sperm cells shown in green, and immotile sperm cells in magenta) and debris (yellow circles), is injected from the left side of the channel, where all the cells and particles follow the fluid flow. In the region of the acoustic actuator, the complete sample is focused inside the middle of the channel. Motile sperm cells escape from the focused flow and are collected via the side channels, while non-progressive/immotile sperm cells and debris follow the flow and are collected via the middle channel. (b-c) (Numerical simulations, see Sec. 2.1) Escape of motile sperm cells (or other motile particles like Janus particles) from the acoustic focusing potential and their separation from immotile species (immotile sperm cells and debris): (b) Strong trapping potential; the motile (the trajectories are shown by green lines) and immotile (red lines) species are focused, and they move through the detector at $x = 5$ while a detector at $x = 15$ detects motile and immotile species separately. (c) Reduced trapping potential to 0.5 of the value in (b): the motile species can partially overcome the trapping potential, and these are detected separately even by the close detector at $x = 5$, as well as by a detector at position $x = 15$. The thermal diffusion parameters: temperature, $T = 2.5 \cdot 10^{-3}$, the rotational diffusion coefficient, $D_{rot} = 2.5 \cdot 10^{-2}$. The unit length is the diameter of the particle.

(a)

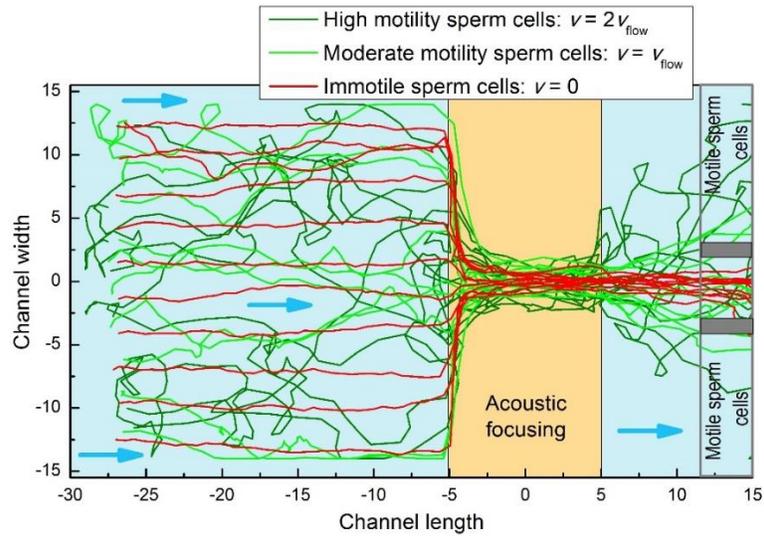

(b)

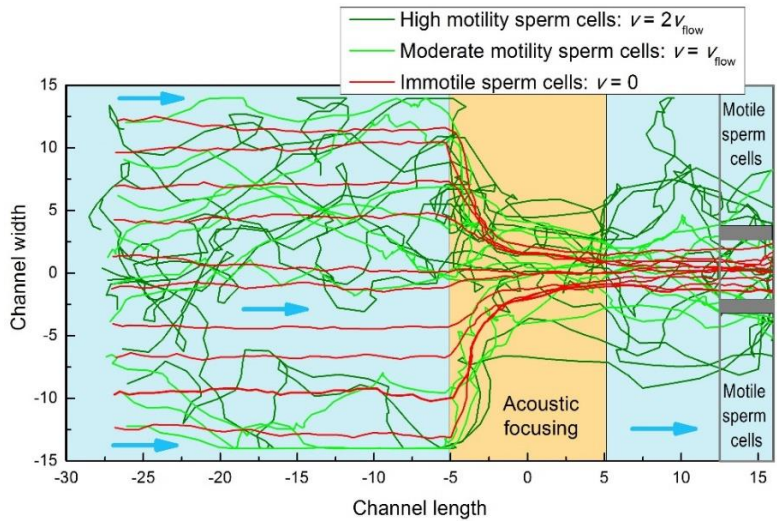

**Figure 2.** (Numerical simulations, see Sec. 2.1) Escape of motile sperm cells (or other motile particles like Janus particles) from the acoustic focusing potential and their separation from immotile species: the motile sperm cells are characterized by high motility (with self-velocity twice that of the fluid flow, $v = 2v_{flow}$), moderate motility ($v = v_{flow}$) and zero motility (immotile sperms and debris). (a) A strong trapping potential that is able to focus all the species. All the species move through the first detector (located at $x = 5$), and they can be separately detected further away from the focusing potential, at $x = 15$. Motile swimmers escape the focused flow and can be detected by the side detectors while immotile species remain focused. (b) A weaker trapping potential: 0.25 of the value shown in (a). The motile sperm cells partially overcome the barrier and thus flow outside the focused flow of non-motile particles. Note that high motility species overcome the barrier easier and thus can be separated not only from immotile species but also from less motile species. In this way, we can select the most motile species such as highly motile sperms needed for successful fertilization.

## 2.2. Experiment: Selection of self-propelled Janus particles

Following the predictions of the numerical simulations of the previous section, we use self-propelled JPs, as a model system for motile sperm, to experimentally verify the predicted behavior. As motile species, JPs are used which are polystyrene (PS) spheres with one hemisphere coated by a 25-nm Pt layer (for fabrication details, see Ref. [40]). These self-propelled particles are fueled by hydrogen peroxide, $H_2O_2$, that is decomposed to water and oxygen at the Pt-coated side of the particle inducing self-propulsion. Immotile (passive) particles are represented by PS beads.

### Infusion of particles and hydrogen peroxide in the chip

The liquids and the sample were infused in a 375-mm wide microfluidic channel (with the depth of 350 mm), with three inlets and either one or three outlets. The continuous flow of the liquids and the particles in the channel was created by the Fluigent pumps via capillaries connected to the inlets. The pressure in the experiments varied between 1 mbar and 15 mbar. To provide the flow velocities to be comparable to the velocity of self-propelled particles, we tuned the pressure down the lowest value of 1 mbar during the measurements of the particle escape from the focused flow, after the initial phase of fast flow (at 20 to 50 mbar) and particle acoustic focusing. The fact that concentration of the active particles (e.g., in the focused flow) was much higher in our experiments than, e.g., in low-density measurements in the absence of external flows [40], required special care. Unlike in the experiments of Ref. [40], where single Janus particles were directly placed in the $H_2O_2$ solution, in our case this would result in a very intensive chemical reaction of decomposition of $H_2O_2$ at the surfaces of Janus particles and consequently in the formation of $O_2$ bubbles (in practice filling the entire microchannel). Therefore, care was taken to avoid any contact of the Janus particles with $H_2O_2$ outside the active zone (where particle escape occurs) in the channel. For this purpose, the sample consisting of Janus particles and passive PS beads in pure water was infused via the central inlet while $H_2O_2$ was infused via the side inlets. This way of infusion guaranteed that the Janus particles interacted with the $H_2O_2$ solution only when flowing inside the channel. Another related precaution was to keep the local density of Janus particles low in the region where they mix up with the side flows with $H_2O_2$ solution. By tuning the concentration of the Janus particles and the flow velocity, we achieved the regime when the interaction of the Janus particles with the $H_2O_2$ solution resulted in a gentle balance, providing the self-propulsion and the same time avoiding the bubble formation. Note that mixing of the side flows ($H_2O_2$) with the central flow (water with Janus particles and PS beads) was provided by the vortical flows across the channel generated by the acoustic actuator attached to the microchannel.

### Focusing the particle flow in the acoustofluidic setup

In our experiments, particles are focused along the centerline of the channel under the action of the acoustic radiation force exerted on the particles from the standing acoustic wave.

The radiation force is a mean force which results from the nonlinear properties of wave motion in fluids [41] and is an analog of the optical radiation force exerted by an electromagnetic wave on electrically or magnetically responsive objects. The theory of acoustic radiation pressure was first proposed by Lord Rayleigh [42]. King [43] derived an expression for the radiation pressure on a rigid sphere, and Yosioka and Kawasima [44] extended his results to compressible spheres. Following the approach of King and Yosioka & Kawasima, several recent studies [45-47] have assumed the surrounding fluid and the medium inside the sphere to be inviscid and non-heat-conducting.

The acoustic radiation force exerted on a compressible, spherical, micrometer-sized particle of radius $a$ suspended in an inviscid fluid in an ultrasound field of wavelength $\lambda$ can be evaluated within the first-order scattering theory (see, e.g., [46]). A small particle, i.e., $a \ll \lambda$, of density $\rho_p$ and compressibility $\kappa_p$ acts as a weak point-scatterer of acoustic waves. An incoming wave described by a velocity potential $\phi_{in}$, results in a scattered wave $\phi_{sc}$ propagating away from the particle. For sufficiently weak incoming and scattered waves, the total first-order acoustic field $\phi_1$ is given by the sum of the two potentials:

$$\phi_1 = \phi_{in} + \phi_{sc}. \tag{1}$$

The corresponding first-order velocity field $\boldsymbol{v}_1$ and pressure $p_1$ are:

$$\boldsymbol{v}_1 = \nabla \phi_1, \; p_1 = i\rho_0 \omega \phi_1, \tag{2}$$

where $\rho_0$ is the fluid density, and $\omega$ is the frequency of the radiation field. The expression for the radiation force, $\boldsymbol{F}^{rad}$, can be presented in the general form [46]:

$$\boldsymbol{F}^{rad} = -\int_{\partial\Omega} da \left\{ \frac{1}{2}[\kappa_0 \langle p_1^2 \rangle - \rho_0 \langle v_1^2 \rangle]\boldsymbol{n} + \rho_0 \langle (\boldsymbol{n} \cdot \boldsymbol{v}_1)\boldsymbol{v}_1 \rangle \right\}, \tag{3}$$

where the integration is performed over any surface $\partial\Omega$ encompassing the sphere, $\kappa_0 = 1/(\rho_0 c_0^2)$ is the compressibility of the fluid, and $c_0$ is the speed of sound in the fluid. After performing the integration in Eq. (3), the resulting radiation force acting on small spherical particle in a stationary acoustic field and inviscid fluid can be presented in a convenient form [41]:

$$\boldsymbol{F}^{rad} = -\left(\frac{\pi p_0^2 V_p \kappa_0}{2\lambda}\right) \phi(\kappa, \rho) \sin\left(\frac{4\pi x}{\lambda}\right), \tag{4}$$

where $\phi(\kappa, \rho)$ is the acoustic contrast factor:

$$\phi(\kappa, \rho) = \frac{5\rho_p - 2\rho_0}{2\rho_p + \rho_0} - \frac{\kappa_p}{\kappa_0}, \tag{5}$$

$p_0$ is the acoustic pressure, and $V_p$ is the volume of the particle.

The acoustic radiation force (4) is linear with respect to the particle volume. Thus, the radiation force decreases rapidly with the particle radius. On the other hand, the competing force exerted on particles under acoustic excitation, is the streaming force due to the rotating fluid (see, e.g., Ref. [23]) that is linear with respect to the particle radius, and it becomes larger than the radiation force for particles with the diameter smaller than ~ 2 μm, for PS or silica particles [23]. In other words, the acoustic focusing can be achieved for PS particles larger than 2 μm (which also depends on the contrast factor, Eq. (5)). We have tested this threshold, by infusing and focusing PS particles with the diameter of 2, 3, 4 and 5 μm. In principle, even 2 μm particles could be focused, but this requires higher acoustic power, while the larger-size particles of 3, 4 and 5 μm have been effectively focused (same as earlier found in Ref. [23] and other related works) even for a moderate acoustic power.

All the infused particles in the channel were initially focused by the radiation force (3). The resonant condition in the channel of 375-μm width was achieved at the excitation frequency of 1.94 ± 0.03 MHz. The voltage applied to the piezo-electric acoustic actuator was tuned between 40 mV (maximum focusing) and 0 (no focusing). First, the voltage was set equal to 40 mV, and then gradually reduced, thus allowing motile particles to escape from the acoustic trap. The results are presented in Figs. 3(a-d). The particles in the middle flow and in the side flows can be collected separately as schematically shown in the plots.

In the experiments with motile Janus particles, we used PS particles with the diameter of 4 μm, covered with a Pt layer. From equations (4) and (5) it is clear, that the acoustic radiation force exerted on such a particle is *larger* than the force acting on an uncovered PS particle of the same size. The difference is due to the acoustic contrast factor (5) which depends on the density of the material of the particle, $\rho_p$, and its compressibility, $\kappa_p = 1/(\rho_p c_p^2)$, (i.e., on the speed of sound in the material, $c_p$). Taking the density of Pt, $\rho_{Pt}$ = 21450 kg/m$^3$, and the speed of sound, $c_{Pt}$ = 2800 m/s, we find that the acoustic contrast of Pt is ~ 4.11 times higher than that of PS (which is relatively low, just slightly higher than that of water). This means that a Pt particle would experience a 4.11 times larger acoustic force than a PS particle of the same size. However, the Janus particles are covered by a thin layer of Pt, and only on one hemisphere of PS particles. The calculation of the contrast factor for such a complex geometry would be rather non-trivial (e.g., it has been calculated in spherically symmetric multi-layer systems [48, 49], revealing non-monotonic dependences of the radiation force on the thickness of the layers and the frequency). Although, even without calculations, it is clear, that adding a portion of a material with a higher acoustic contrast factor would result in an *increase* of the acoustic factor of the complex particle. Therefore, if we would like two different species of particles with different contrast factor $\phi(\kappa, \rho)$ (5) to experience *the same* acoustic radiation force (4), the particles with a smaller $\phi(\kappa, \rho)$ should be of a *larger* size. In case of the Janus particles, the increase in the contrast factor $\phi(\kappa, \rho)$ should be compensated by an increase in size of PS particles. Following these considerations, we have chosen slightly larger passive beads, i.e., PS particles with the diameter of 5 μm.

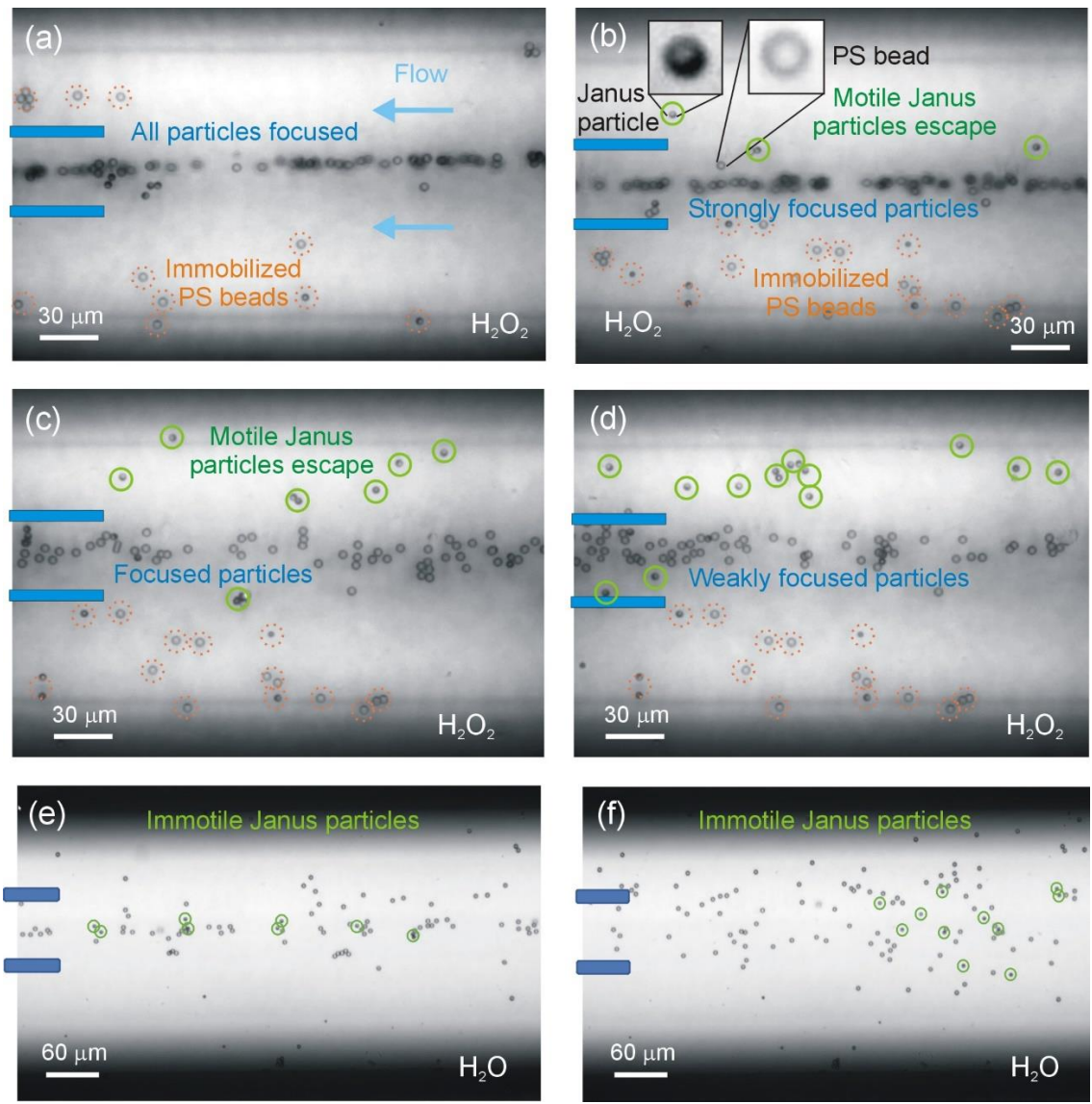

**Figure 3**. (Experiment) Escape of motile (a-d) Janus particles (the Pt-covered part is dark, and the PS open part is bright, see insets in (b)) from the acoustic focusing potential and their separation from non-motile species (passive beads, bright particles). The flow is from the right to the left. The blue rectangles on the left show the gate separating the central flow from the side flows. (a) The focusing potential is strong (U = 40 mV): all the particles, active and passive, are focused inside the straight narrow band and are collected inside the gate; no separation. (b) The focusing potential is weaker (U = 20 mV): a few of the Janus particles escape the focusing potential; the onset of separation. (c) U = 12 mV: at a weaker focusing potential, most of the Janus particles are detected outside the gate, i.e., separated from the rest of the particles. Although the focused band of passive particles becomes broader (cp. (b)), all the passive particles are collected inside the central gate. (d) Same as in (c), after 3 min. The picture remains the same (steady state) as in (c): motile Janus particles are separated from the passive beads. (e-f) Flow of a mixture of Janus particles (marked by the green circles) and passive beads in pure water (i.e., in the absence of chemical activity of Janus particles), in the presence of the acoustic focusing potential. (e) U = 10 mV: the flow is weakly focused; some of the passive particles escape the flow while the Janus particles remain focused showing no tendency of escape. (f) U = 2 mV: the flow of the particles is substantially broadened, and the distribution of the Janus particles does not show any preferential escape from the flow as compared to the passive beads.

Since this compensation is not exact, we performed an additional control experiment in pure water confirming that the escape of Janus particles from the focused flow is indeed due to their motility, and not due to any other potentially competing effects. These effects are not only due to the difference in size of JPs and PS particles but could also be related to a torque [50] or even self-propulsion induced by the microwaves [51-53]. To examine whether those effects could lead to any selectivity, the additional control experiment was performed in pure water, excluding chemical activity. The experiment revealed no sign of selective escape from the focusing flow, as shown in Fig. 3(e-f) thus proving that the selective escape of Janus particles from the focusing flow is essentially due to motility of the active swimmers. This confirms that we observe separation of particles based on their motility using the proposed acoustofluidic setup.

**2.3. Experiment: Selection of human sperm cells based on their motility**

The experiments were conducted with an already processed semen sample using density gradient centrifugation, with concentration of 60 million sperm cells per ml with about half of the cells motile (45% grade A+B and 3% grade C) and the other half immotile (52% grade D).

In comparison to $H_2O_2$ activated Pt-covered Janus particles, sperm cells have much higher self-velocity of the order of ~ 60 to 80 mm/s. The advantage of the high propulsion velocity is that it enables the realization of the optimal balance between the self-propulsion velocity and the flow velocity for the effective separation on motility, without the necessity of exploring the regime of very low flow rates when the flow could become unstable. Considering these factors, and taking the necessary precautions (i.e., the temperature control at high acoustic power), the experiments with human sperm cells have been carried out to reveal the predicted selection according to their motility.

Human sperm cells are rather sensitive to the heating of the chip with the acoustic actuator. The optimal temperature range for motile sperm cells is between room temperature and 37°C. To avoid the effect of heating, we employed our developed cooling system that includes a thermoelectric Peltier element, one side of which is attached to the chip, and the other side to an external cooling circuit with cold water that removes the excess of heat from the chip. The range of the operating temperatures has been controlled via an electronic circuit connected to the Peltier element. We note however that for relatively low acoustic power (corresponding to the applied voltage of up to 50 mV) the heating of the active area of the chip (monitored by an IR temperature-control camera) did not exceed 37°C and thus was safe for sperm cells.

Furthermore, the method we employed does not imply a contact of sperm cells with mechanical obstacles (like guiding lateral walls or corners, unlike, e.g., in passive-driven devices) in the focusing (separation) zone, which significantly reduces the danger of sperm cell damage as well as undesired jamming. Also, we demonstrated that overall, our acoustofluidic setup is safe for sperm cells (including all potential harm factors), which was confirmed in the viability test: the motility of cells that passed via the setup in the presence or in the absence of the acoustic signal (of the maximum power employed in the experiments) was compared with the motility of control samples and showed no correlation or any appreciable difference.

The flow patterns of the sperm sample in the presence of the acoustic focusing potential for U = 50 mV, are shown in Fig. 4. Due to the high velocity of motile sperm, low motility (referred to as grade

C) and immotile (grade D) sperm cells appear to be trapped by the focusing potential inside a narrow band around the centerline of the channel, while only highly progressive motile sperm cells *freely move* inside the entire channel and are present outside the focused centerline band (see Fig. 4(b)).

These highly motile cells can then be collected separately via the side outlets, as schematically shown in Fig. 4. The rest of the cells, including immotile and non-progressive motile sperm, leave the channel via the central outlet. The selection criterion can be adjusted by changing the width of the central outlet. Thus, Design I schematically presented in Fig. 4(b) is suitable for the selection of progressive motile sperm cells with high velocity from progressive sperm with lower velocity, non-progressive motile and immotile sperm cells, i.e., when one needs to select a small amount of the *highest motility* sperm. On the contrary, Design II with a narrower central outlet enables selecting immotile cells from progressive motile sperm cells with high and low velocity. This design can be used when we only need to *exclude immotile* sperm cells from the sample.

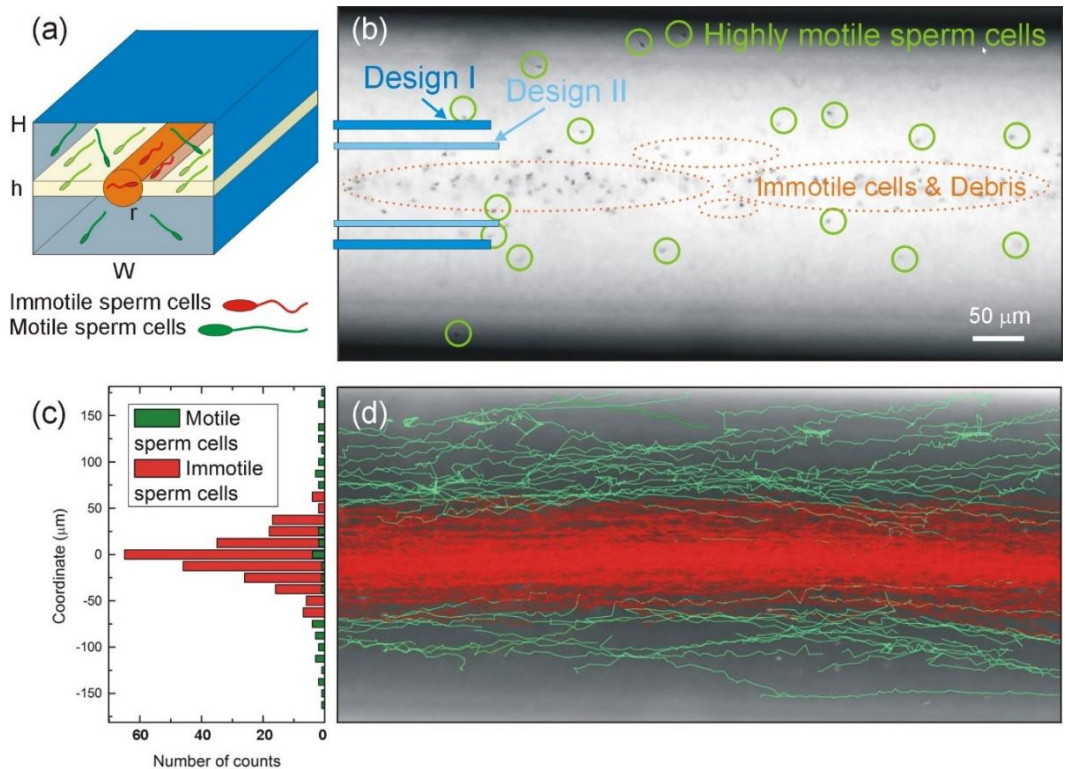

**Figure 4**. (Experiment) (a) A sketch showing the distribution of motile and immotile sperm cells inside the channel and in the visible area. (b) A snapshot of the flow of a mixture of highly motile sperm cells (indicated by the green circles) and low-motility and immotile sperm cells (indicated by the dark orange ellipses), in the presence of the acoustic focusing potential, U = 50 mV. The immotile and low-motility sperm cells are trapped by the focusing potential along the centerline of the channel. Highly motile sperm cells escape the trap and freely move inside the entire width of the channel. Thus, only highly motile cells are present outside the centerline. These motile cells can be separately collected via the side outlets (indicated by the thick blue lines) while all the immotile cells are collected via the central outlet. Design I: a wider central outlet enabling only progressive motile sperm cells with high velocity to be collected at the side outlets. Design II: a narrower central outlet enabling progressive motile sperm cells with high and low velocity to be collected at the side outlets, while only immotile cells move via the central outlet. (c) Count histogram of progressive motile (e.g., grade A+B, green) and immotile sperm (e.g., grade C+D, red). (d) Tracked trajectories of motile (green) and immotile (red) sperm cells moving in the channel. The trajectories of immotile cells are straight, and the trajectories of motile cells show steps and the motion across the streamlines.

**Counting motile and immotile sperm cells in the field of view of the microscope**

Let us analyze the results of the observations presented in Figs. 4(b) and (d). The numbers of progressive motile and immotile sperm cells infused in the channel are nearly equal, and one can expect to observe (assuming the sorting capacity of the device to be equal to 1) the same ratio in the output count. This would be indeed the case if we were able to count all the sperm cells in the entire volume of the channel. However, only a thin quasi-2D layer of thickness $h$ (within the focal depth) is observed via the microscope, as illustrated in Fig. 4(a).

It is clear then, that in this way in the field of view of the microscope we can observe only a *small fraction* of motile sperm cells. On the contrary, almost *all immotile* sperm cells appear to be *within* the focus depth (which are focused inside the cylinder near the center of the channel as shown in Fig. 4(a)). Simple geometric estimates result in the following ratio of the number of cells outside the focused flow (motile cells), $N_{m,out}{}^{3D}$, to the number of cells inside the focused flow formed mainly by immotile cells and fewer motile cells, ($N_{m,in}{}^{2D} + N_{im,in}{}^{2D}$), observed within the quasi-2D layer with the focal depth $h$:

$$\frac{N_{m,out}^{3D}}{N_{m,in}^{2D}+N_{im,in}^{2D}} = \frac{N_m h(W-2r)\pi r}{2h(N_{im}WH+2N_m\pi r^2)}, \tag{6}$$

where indices $m$ and $im$ refer to motile and immotile cells, $h$ is the depth of the observed layer (in focus of the microscope), $W$ and $H$ are the width and the height of the channel, and $r$ is the radius of the cylindrical area of the focused flow (see Fig. 4(a)).

Taking, e.g., $W$ = 375 μm, $H$ = 350 μm, $r$ = 35 μm and $N_m \approx N_{im}$, we obtain:

$$\frac{N_{m,out}^{3D}}{N_{m,in}^{2D}+N_{im,in}^{2D}} \approx 0.12. \tag{7}$$

Therefore, for equal initial numbers of motile to immotile cells and the sorting efficiency $C_s$ = 1, the ratio of the motile sperm cells in the field of view of the microscope to that of immotile cells is about 0.12. Also, out of all the cells inside the focused area, nearly 97% are immotile cells and 3% are motile cells. Therefore, the ratio of the number of motile sperm cells counted outside the focused area to that of immotile cells inside the focused area is:

$$\frac{1}{\eta} = \frac{N_{m,out}^{3D}}{N_{im,in}^{2D}} \approx 0.124. \tag{8}$$

Here $1/\eta$ is the correction coefficient equal to ratio of the number of cells observed in the quasi-2D layer to the total number of cells in 3D. Therefore, the sorting capacity of the method (based on the counts of cells via the microscope), $C_s$, can be defined as follows ($\eta_0$ stands for the initial ratio of motile to immotile sperm cells in the sample):

$$C_s = (N_{motile}/N_{immotile})\eta/\eta_0. \tag{9}$$

In the counting diagram shown in Fig. 4(c), the counts of motile and immotile cells are 26 (outside the focused are) and 253 (inside the focused area), correspondingly, that results in the sorting efficiency (for the particular measurement, for $\eta_0$ = 48/52) $C_s \approx 0.90$.

## 3. Conclusions

We demonstrated the principle of separation of human sperm cells or other motile self-propelled particles according to their motility, in an acoustofluidic setup. The separation is based on the ability of motile species to escape from the trapping potential created by the acoustic radiation force on the nodes of the acoustic standing waves. Immotile species remain trapped and, in this way, can be separated from the motile species and collected separately.

The proposed principle has been realized in "numerical experiments" performed using Molecular Dynamics simulations of self-propelled particles in the presence of a tunable trapping potential, and subsequently verified in the experiments with self-propelled Janus particles, as a model system for motile sperm cells, and ultimately in the experiments with motile and immotile human sperm cells.

Exploring the flexibility of the "numerical experiments", we revealed that for the optimal separation, a balance is required between the key parameters of the system: (i) the self-velocity of the motile sperm, (ii) the flow velocity, and (iii) the strength of the trapping potential. Thus, it was shown that the flow velocity, on the one hand, should be high enough to support the directed transport of the motile and immotile species to the outlet channels. On the other hand, the flow velocity should not be too high, as compared to the self-velocity of motile sperm, to enable the escape of motile species and their displacement sufficiently far away from the focused flow, to be collected separately. The geometry of the setup also plays an important role for the separation effect. The strength of the trapping potential should be sufficient to trap at least all immotile species (and eventually non-progressive motility species or progressive species with lower velocity) and allow motile species with the highest velocity to escape the central channel. By tuning the trapping potential, it is possible to separate motile sperm cells with different velocities. This can also be achieved by a proper choice of the width of the central outlet, as has been shown in the experiments with sperm cells.

The separation based on motility, demonstrated in the numerical simulations, has been verified in the experiments with self-propelled Pt-coated Janus particles activated by hydrogen peroxide. As immotile species, passive PS beads were used. First, motility of the Janus particles was evaluated. The self-propulsion velocity was found to be 1 to 2 $\mu$m/s. Correspondingly, the flow velocity was tuned to the minimum possible values of about 2 to 10 $\mu$m/s. This allowed us to observe the escape of motile Janus particles from the acoustic focusing trap at about 10 to 15 mV. In this way, motile Janus particles were selected from the PS beads. The selected species were transported to the end of the channel where they can be collected separately. To ensure that the escape is essentially due to the motility of the Janus particles (and not due to other potential effects related to the acoustic excitation of Janus particles and PS beads), a control experiment was conducted using the same sample at the same conditions, but without hydrogen peroxide. The control experiment in pure water showed no separation, thus, supporting the conclusion that the observed separation is indeed due to the motility.

Finally, the separation according to motility has been demonstrated in experiments with processed human sperm. The experiments were conducted using a sample containing nearly equal concentration of motile and immotile sperm cells and the optimal flow velocities (commensurate with the motile sperm self-velocity of about 60 to 80 $\mu$m/s). The setup was equipped with a state-of-the-art cooling system that allowed us to avoid possible overheating. In the experiments, the escape of motile sperm cells was observed from the focusing trap at U = 50 mV, which was strong enough to trap all the immotile sperm cells. In this way, we achieved the efficient separation: all immotile species were focused inside the central narrow band near the centerline of the channel, and only motile sperm cells were present outside the narrow central band. We showed that the motile and immotile species can be collected separately at the end of the channel, via separate outlets. Two designs of the central outlet were proposed: wide and narrow. The wider outlet allows to separate only progressive motile sperm cells with the highest velocity. By using a narrower central outlet, all immotile sperm cells will be removed.

We emphasize that the method presented in this work provides much more flexibility in selecting sperm cells based on their motility than passive methods. This is achieved thanks to the tunability of the acoustic trap that can localize the motion and separate motile from immotile sperm cells. Such a flexibility, provided by our method, was not achievable when using a passively driven sperm separator that is the key element of many sperm separation techniques proposed so far.

Furthermore, our method is non-contact, that significantly reduces the danger of sperm cell damage. Finally, sperm motility is not affected by our acoustofluidic setup, as compared to the control samples as shown in the viability test (i.e., evaluating the motility of sperm cells before and after the experiment).

To conclude, the separation of sperm cells according to their motility has been demonstrated in experiments with human sperm as well as in numerical simulations and in experiments with a model system, i.e., self-propelled Janus particles. This study opens new perspectives for the elaboration of efficient continuous methods of motile sperm separation for assisted reproductive techniques and poses new challenges to be addressed, such as selection of sperm cells with low DNA fragmentation. Further efforts will be devoted to the implementation of the elaborated designs and their optimization for the best performance of the proposed separation technique and building a fully operational device for improved assisted fertilization. In addition, in our future studies, we plan to perform experiments with high-motility visible light-driven Ag/AgCl Janus micromotors that propel in pure water [22, 35], i.e., in bio-compatible environments.

## 4. Materials and methods

### 4.1. Numerical methods: Molecular-dynamics simulations

The behavior of the system consisting of motile microswimmers (artificial self-propelled Janus particles or sperm cells) and passive species (synthetic beads or immotile sperm cells and debris) is simulated by numerically integrating the overdamped Langevin equations [16, 21, 22, 32, 33, 35-37, 39-40]:

$$\dot{x}_i = v_0 \cos\theta_i + \xi_{i0,x}(t) + \sum_{ij}^N f_{ij,x} + v_{\text{flow},x} + f_{r,x},$$
$$\dot{y}_i = v_0 \sin\theta_i + \xi_{i0,y}(t) + \sum_{ij}^N f_{ij,y} + v_{\text{flow},y} + f_{r,y}, \quad (10)$$
$$\dot{\theta}_i = \xi_{i\theta}(t),$$

for $i, j$ running from 1 to the total number $N$ of particles, active and passive; $v_0$ is self-velocity of active particles. Here, $\xi_{i0}(t) = (\xi_{i0,x}(t), \xi_{i0,y}(t))$ is a 2D thermal Gaussian noise with correlation functions $\langle\xi_{0,\alpha}(t)\rangle = 0$, $\langle\xi_{0,\alpha}(t)\xi_{0,\beta}(t)\rangle = 2D_T\delta_{\alpha\beta}\delta(t)$, where $\alpha, \beta = x, y$ and $D_T$ is the translational diffusion constant of a passive particle at fixed temperature; $\xi_\theta(t)$ is an independent 1D Gaussian noise with correlation functions $\langle\xi_\theta(t)\rangle = 0$ and $\langle\xi_\theta(t)\xi_\theta(0)\rangle = 2D_R\delta(t)$ that models the fluctuations of the propulsion angle $\theta$.

Since we were interested in establishing relations between the parameters for the optimal selection, the numerical simulations were performed here for the dimensionless units: the unit length is the diameter of the particle, $2r_0$; the damping coefficient, $\gamma$, is set to unity; the unit force (and velocity) is $v_{\text{flow}}$; the thermal diffusion parameters: temperature, $T = 2.5\cdot10^{-3}$, the rotational diffusion coefficient, $D_{\text{rot}} = 2.5\cdot10^{-2}$. This makes the found relations universal and applicable for various motile particles or cells. If needed, these can be easily converted in specific (dimensional) parameters for a particular system. For example, for Janus particles, the diffusion coefficients, $D_T$ and $D_R$, can be directly calculated or extracted from experimentally measured trajectories and mean squared displacement curves (MSD), by fitting to theoretical MSD [35]. Thus, for a particle with diameter of 2 μm diffusing in water at room temperature, $D_T \approx 0.22$ μm$^2$ s$^{-1}$ and $D_R \approx 0.16$ rad$^2$ s$^{-1}$.

The term, $\sum_{ij}^N f_{ij}$, represents, in a compact form, the inter-particle interaction forces in the system, i.e., elastic soft-core repulsive interactions between active particles, between passive beads, and between active and passive particles (for simplicity, it is common (see, e.g., [16, 32]) to present both species by soft interacting disks of radius $r_0$ and repulsive force of modulus $F_{i,j} = k(2r_0 - r_{ij})$ if $r_{ij} < 2r_0$ and $F_{i,j} = 0$ otherwise, where $k$ is the interaction constant); $v_{\text{flow}}$ is the flow velocity, and $f_r$ is the acoustic radiation force exerted on the particle. Note that the proportionality coefficient between velocity and force in the overdamped equations (10), i.e., the cumulative damping constant $\gamma$, is set to unity. It is also assumed that in the overdamped regime the driving force due to the flow is balanced by the Stokes drag, and the net component of velocity of a particle is equal to the fluid flow. The radiation force profile along the *y*-direction, i.e., across the channel, changes as $A_r \cdot \cos(y)$, with a minimum at the centerline of the channel, where $A_r$ is the amplitude of the radiation force. It is also set to smoothly decrease as ~ $\cos(x)$ from the current value (for certain *y*) to zero along the *x*-direction at the boundaries of the action of the acoustic excitation.

It is worth noting that the sorting mechanism is only efficient in the absence of (strong enough) torques that would align the motile species along the trap direction (i.e., along the low direction). Janus particles, as discussed in Sec. 2, tend to move across the streamlines, i.e., away from the acoustic trap, and in this way escape. In case of sperm cells, the radiation force is exerted on the head of the sperm (that is typically 2 to 5 μm in size) while the narrow tail (less than 2 μm in diameter) is unaffected. This allows sperm cells to move cross-stream (unlike in a narrow channel where the motion of sperm would be aligned allowing no selection).

**4.2. Experimental methods: Acoustofluidic setup**

The acoustofluidic setup for the experiments with particles consists of a Leica DMi8 microscope, an inverted light microscope, that is used as a general-purpose laboratory microscope for routine examination of biological samples. The inverted microscope is equipped with a CCD fluorescent camera (C13440-20C, Hamamatsu Photonics, Japan). A piezoceramic element (PZT) measuring 15 mm x 20 mm x 1 mm (APC International Ltd., USA) is connected to the microfluidic chip and positioned above the channel generating bulk acoustic waves. The PZT is coupled to the microfluidic chip by a thin glycerol layer. An in-house built PMMA holder ensured a good connection between the actuator and the chip. The PZT was driven by a frequency generator (Keysight 33500B series, USA). The amplifier N-DP 90 Prana with an output power of 90 Watt has been used to increase the applied voltage. At high power, corresponding to a voltage of more than 100 mV, the PZT and the microfluidic chip are subjected to excessive heating, which could be harmful to living organisms, such as bacteria and sperm cells. Therefore, the setup is equipped with a cooling system and a PID controller, which is used in experiments with sperm cells. Otherwise, the heating is insignificant with voltages up to 40 mV and synthetic particles.

The microfluidic chip used in the experiments with Janus particles and passive beads contains a main channel with a height of 350 μm and a width of 375 μm, with a theoretical resonance frequency of 2.00 MHz. The length of this channel is 53 mm. The device has three inlet silica capillaries and one outlet capillary. The outer (inner) diameter of these capillaries is 350 μm (200 μm) (TSP200350, CM Scientific, United Kingdom). The capillaries were glued with UV glue (Ormocore from microresist technology GmbH, Germany). The microfluidic chip was fabricated in the MESA+ clean room (UTwente, The Netherlands). The channels were patterned with a positive photoresist (Olin 907-12) using mid-UV lithography. They were then etched into a silicon wafer with a thickness of 525 μm using Bosch-type Deep Reactive-Ion Etching (DRIE) (Adixen AMS100SE, Pfeiffer Vacuum SAS, France). Afterwards, the resist was removed with nitric acid and oxygen plasma. The microfluidic channels were finally sealed with a 500 μm thick Pyrex wafer by anodic bonding. An identical chip has been used in the experiments with human sperm cells.

The oscillating frequency of the PZT has been adjusted via the frequency generator. The actual resonance frequency providing the best focusing (narrowest band) of the sample can slightly vary, depending on the concentration of hydrogen peroxide and on the injected sample and was typically within 1.94 ± 0.03 MHz.

**4.3. Use of human sperm cells**

Approval was obtained from the local ethical committee of UZ Brussel to perform experiments on human sperm (BUN 1432022000168). All the methods used in this study comply with the relevant guidelines and regulations including obtaining the corresponding informed consent from the sperm donors when relevant. Sperm cells were analyzed by the qualified designated personnel at the Center for Reproductive Medicine, UZ Brussel.

## Author Contributions

Conceptualization: V.R.M. and W.D.M. Data curation: V.R.M. Formal Analysis: V.R.M. Funding acquisition: W.D.M. and D.M. Investigation: V.R.M., with the contribution from W.D.M., T.H., P.G., I.M., K.W., N.D.M., and F.N. Methodology: V.R.M., W.D.M., P.G., L.B., D.M., T.H., I.M., K.W., N.D.M., and F.N. Project administration: W.D.M. Resources: W.D.M. Software: V.R.M. and W.D.M. Supervision: W.D.M., D.M., and N.D.M. Validation: V.R.M. Visualization: V.R.M. and T.H. Writing – original draft: V.R.M. and W.D.M. Writing – review & editing: all.

## Acknowledgements

V.R.M., D.M. and W.D.M. acknowledge the support of Research Foundation-Flanders (FWO-Vl), Grant No. G029322N. F.N. is supported in part by: Nippon Telegraph and Telephone Corporation (NTT) Research, the Japan Science and Technology Agency (JST) [via the Quantum Leap Flagship Program (Q-LEAP) and the Moonshot R&D Grant Number JPMJMS2061], the Asian Office of Aerospace Research and Development (AOARD) (via Grant No. FA2386-20-1-4069), and the Foundational Questions Institute Fund (FQXi) via Grant No. FQXi-IAF19-06.
The authors thank Antonio Maisto for the help with the use of the acoustic setup and the cooling system.